Article

# Remote-Controlled Activation of the Release through Drug-Loaded Magnetic Electrospun Fibers

Richard Ziegler [1], Shaista Ilyas [1], Sanjay Mathur [1], Gerardo F. Goya [2,3] and Jesús Antonio Fuentes-García [2,*]

[1] Institute of Inorganic Chemistry, University of Cologne, Greinstr. 6, 50939 Cologne, Germany; richie.ziegler@arcor.de (R.Z.); silyas@uni-koeln.de (S.I.); sanjay.mathur@uni-koeln.de (S.M.)
[2] Instituto de Nanociencia y Materiales de Aragón (INMA), CSIC-Universidad de Zaragoza, Campus Río Ebro, 50018 Zaragoza, Spain; goya@unizar.es
[3] Departamento de Física de la Materia Condensada, Facultad de Ciencias, Universidad de Zaragoza, 50009 Zaragoza, Spain
* Correspondence: j.fuentes@unizar.es

**Abstract:** The integration of magnetic nanoparticles within fibrillar structures represents an interesting avenue for the remotely controlled release of therapeutic agents. This work presents a novel drug release platform based on electrospun magnetic fibers (EMFs) combining drugs, magnetic nanoparticles (MNPs) and mesoporous silica nanoparticles (MSNs) for controlled drug delivery via alternating magnetic fields (AMF). The platform was demonstrated to be versatile and effective for hydrophilic ketorolac (KET) and hydrophobic curcumin (CUR) encapsulation and the major response observed for AMF-triggered release was reached using drug-loaded MSNs within the fibers, providing fine control over drug release patterns. The EMFs exhibited excellent inductive heating capabilities, showing a temperature increase of $\Delta T$ up to 8 °C within a 5 min AMF pulse. The system is shown to be promising for applications like transdermal pain management, oncological drug delivery, tissue engineering, and wound healing, enabling precise control over drug release in both spatial and temporal dimensions. The findings of this study offer valuable insights into the development of the next generation of smart drug delivery systems, based in multifunctional materials that can be remotely regulated and potentially revolutionize the field of nanomedicine.

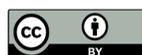

## 1. Introduction

Electrospun fiber-based drug delivery systems have demonstrated tremendous potential in facilitating the controlled release of therapeutic molecules [1–3]. However, there is a need for a wider range of therapeutic materials utilizing electrospun magnetic fibers (EMFs), particularly for externally triggered drug delivery [4–7]. EMFs can be developed to encapsulate drugs or therapeutic agents within their structure, acquiring magnetic responsiveness through the incorporation of magnetic nanoparticles (MNPs) [8–11].

When AMFs are applied to the embedded MNPs, several phenomena occur, enabling remote-controlled release. These include magnetic heating [12], mechanical actuation [13], spatial control [14], and the modulation of the release rate [15]. Concerning the capacity for remote magnetic heating, the efficiency with which magnetic nanoparticles (MNPs) respond to alternating magnetic fields (AMFs) is strongly influenced by their size, magnetic anisotropy, and degree of agglomeration. Therefore, the selection and engineering of the magnetic material are crucial for attaining effective heating under a magnetic field. Notably, recent studies have demonstrated that the AMF-induced behavior of electrospun magnetic fibers (EMFs) also contributes to enhanced drug release kinetics.

The implementation of EMFs offers several advantages and opportunities to improve the precision, control, versatility, and efficiency of drug release [16–19]. Varying the strength, frequency, or duration of the exposure to AMF, the release kinetics of the drug can be modulated, enabling tailored dosing schedules and the ability to deliver drugs at specific time points. The use of a trigger that interacts non-invasively and non-harmfully with biological barriers, such as tissues or organs, allows for magnetically remote-controlled liberation from EMFs. This on-demand release mechanism holds great promise as a strategy to overcome limitations associated with conventional drug delivery systems, such as poor spatial precision and lack of temporal control.

Magnetically triggered drug release is advantageous in settings where physical intervention is difficult or unwanted, and provides EMFs with an expanded range of applications [20]. Additionally, the incorporation of MNPs enables EMFs to serve multifunctional roles with theranostic capabilities integrated into a single system, for instance, MRI detection and controlled drug release [21]. This integration allows for the simultaneous monitoring of drug release and imaging of targeted areas, enhancing the precision and effectiveness of treatment.

Sonochemical synthesis methods, through the application of ultrasound in a liquid medium, offer advantages such as rapid synthesis, scalability, and low environmental impact [22,23]. The method enables the control of the final size, composition and magnetic properties of the MNPs required in magnetically driven inductive heating [24]. By generating grams of MNPs in just a few minutes, these methods provide a practical and efficient way to obtain the necessary amount of magnetic material for effective heating applications.

Electrospinning is a versatile technique that enables the fabrication of fibers with high surface area-to-volume ratios and tunable properties [25]. In the case of polymeric fibers, it is possible to elaborate the precursor solutions by dissolving the polymer in an appropriate solvent or by melting it. During electrospinning, a high voltage is applied to a precursor solution or melt that is dispensed through a spinneret. The high voltage provides enough electrical force to overcome the surface tension of the viscous liquid, forming Taylor's cone. A jet of the polymer solution or melt is then emitted from the apex of the cone, resulting in the formation of electrospun fibers [26].

Since EMFs allows for the incorporation of various molecules and nanoparticles within the polymer solution, this capability provides opportunities for tailoring the functionality and properties of the electrospun mats according to specific requirements [27,28]. The selection of appropriate polymer precursors is crucial, as their viscoelastic properties directly impact the electrospinning process. Precursor solutions with suitable rheological behaviors, such as proper viscosity, elasticity, and surface tension, are necessary to ensure successful fiber formation. Choosing the right polymer is essential for achieving the desired properties and responses of the electrospun mats [29–34].

In this work, we have developed a novel platform for the transdermal and localized delivery of both soluble and non-soluble drugs, magnetically triggered to control the release and to apply local heating. The use of the biologically insoluble matrix of polyacrylonitrile (PAN) aimed to maintain the integrity of the electrospun mats and prevented the unwanted leaching or dissolution of the polymer and the embedded MNPs. While the PAN polymeric base provides mechanical strength and chemical stability, the addition of MNPs enables the magnetic control of the material temperature and thus releases the dosage at the correct time.

## 2. Materials and Methods

*2.1. Materials*

Curcumin (CUR, from Curcuma longa (Turmeric)), ketorolac tromethamine (KET, Pharmaceutical Secondary Standard; Certified Reference Material), N,N-Dimethylforma-mide (DMF, anhydrous, 99.8%), and Polyacrylonitrile (PAN, average Mw 150,000 (Typical)) were purchased from Sigma-Aldrich (Darmstadt, Germany). Acetate (pH 5.5) and phosphate (pH 7.4) buffers were prepared in deionized water (Millipore 18.2 MΩ·cm @ 25 °C).

MNPs were elaborated using sonochemistry and MSNs using wet chemistry methods, described in Materials Elaboration section of the SI file. MNPs with average size of 21 ± 4 nm; $Fe_{2.9}Mn_{0.1}O_4$ composition; magnetization saturation (Ms) 54 emu/g; coercive field (Hc) 12 Oe; and specific loss power (SLP) 85 W/g at 765 kHz and 364 G were employed. MSNs with average diameter of 33.3 ± 4.2 nm; 845.3 $m^2$/g surface area; 0.97 $cm^3$/g pore density; and 5.58 nm pore size, were loaded with 1 μmol of CUR and 0.38 μmol of KET, reaching maximum release (37 °C) of 2.2% after 12 days at pH 5.5 and 1.2% at pH 7.4 for CUR. While for KET, it reached maximum release of 11.4% after 12 days at pH 7.4 and 8.7% at pH 5.5. Complete dataset of characterization, properties and release profiles from the employed nanoparticle can be consulted in the Supplementary Materials file.

*2.2. Electrospun Magnetic Fibers (EMFs) Processing*

The drug-loaded EMFs electrospinning strategy was developed by mixing all the components in the precursor solutions, providing a simple processing route. Four different electrospinning solutions were prepared to produce four different drug-loaded EMFs samples. All of them consist of a basis of 83.3 mg of MNPs and 500 mg of PAN dissolved in 5 mL of DMF. Additionally, either 4.9 mg of KET, 18.4 mg of CUR, 109 mg KET-MSNs or 139 mg CUR-MSNs were added to the solutions to produce the EMFs-KET, EMFs-CUR, EMFs-MSNs@KET, and EMFs-MSNs@CUR samples. For the precursor solution's elaboration, DMF was mixed with the components of the solution, except PAN, and heated up to 80 °C. After 30 min of magnetic stirring and heating, PAN was added. The stirring and heating continued for 2 h. After this time, there was heating in power-down and the solution continued stirring over-night.

The electrospun polymeric fiber-based magnetic films were processed for 4 h each, injecting the previously prepared precursor solutions in a laboratory-made version of the classical electrospinning system. It consists of a LEGATO® 101 syringe pump (Kd Scientific, Holliston, MA, USA) at 0.7 mL/h for controlled injection placed at working distance (distance needle-collector) of 15 cm from a rotative collector. The 8 kV employed during the processing was provided using a Power Supply Unit (Genvolt®, 30 W, 200 V–30 kV) connected to the needle in the positive pole, while the negative pole was connected to the metallic collector covered with aluminum foil.

*2.3. Electrospun Magnetic Fibers (EMFs) Characterization*

Physical and chemical properties from the EMFs were determined using different techniques. For the morphology and elemental composition, Scanning Electron Microscopy (SEM) was used in a FEI (Hillsboro, OR, USA) INSPECT 50 equipped with field emission gun of single crystal tungsten. The samples were prepared via fixing on holder trimmed mats with double-face carbon film and evaporating carbon coating. Secondary electron images were analyzed for the average diameter determination using Image J software (1.54 g version). For the elemental quantification, energy dispersive X-ray spectroscopy (EDS) mode was employed, recording the spectrum and mapping images using the INCA PentaFETx3 detector from Oxford Instruments (Abingdon, Oxfordshire, UK). Both observations were performed at 30 kV. As-prepared EMFs were analyzed using Fourier Transform Infrared spectroscopy (FTIR) on a Perkin Elmer (Shelton, CT, USA) 1000 spectrometer in the 4000 to 600 $cm^{-1}$ intervals for the functional group's identification. Also, Thermogravimetric Analysis (TGA) was performed in a Mettler Toledo (Columbus, OH, USA) AG—TGA/SDTA851e from 50 to 600 °C at 10 °C/min. Magnetization measurements M(H) were carried out on a SQUID magnetometer (MPMSXL Quantum Design, Barcelona, Spain), measuring the EMFs encapsulated in gelatin capsules. Specific loss power (SLP) measurements were carried out in the AMF applicator nB nanoscale Biomagnetics (Spain) D5 Series, detecting the rising temperature with optical fiber in 1 mL of water using 3 mg of EMFs.

*2.4. Electrospun Magnetic Fibers (EMFs) Release Evaluation*

For the in vitro drug release evaluation of the different drug-loaded EMFs samples, phosphate-buffered saline with pH 7.4 to simulate the physiological environment of blood and acetate buffer at pH 5.5 to simulate the more acidic pH value in tumor tissues was employed. The samples were incubated in a shaker for up to 10 days at 37 °C, with the supernatant being removed and replaced with fresh buffer in a time window of 1–240 h. These aliquots were then tested via UV-Vis spectroscopy and compared to the calibration curves to evaluate their drug concentration and thus the drug release profile of the polymer scaffolds.

On the other hand, for the drug release with applied magnetic field, the samples were prepared as mentioned previously, but in this case, instead of incubating, they were placed within a glass vial into the liquid carrier for magnetic field stimulation. A magnetic applicator (765 kHz at 364 G) was programmed for intervals of 5 min field on, then stopped for aliquot sampling, removing the supernatant for UV-Vis testing and replacing with fresh buffer. The release was tested during cycles for up to 150 min.

**3. Results and Discussion**

Despite using a strategy based on PAN as a gatekeeper for this drug release platform and incorporating MNPs, MSNs and drugs in the same system could represent different challenges in the materials processing; for instance, through ensuring a continuous and homogeneous fibrillar configuration, keeping the components within the EMFs, and preserving adequate magnetic properties for inductive heating, the robustness of the electrospinning technique allows the control of the parameters, obtaining the adequate values to produce homogenous fiber mats with an area of around 400 cm$^2$ (Supplementary Materials, Figure S7).

*3.1. Materials Properties*

All the samples were hydrophilic, except for EMFs-CUR, which shows hydrophobic behavior with a contact angle of 143.6° (Figure S6), according to the contact angle measurement described in Section S2.2 in Supplementary Materials. This effect can be associated with the higher contents of non-soluble CUR in this sample. As evidence of the successful elaboration of EMFs with controlled shape and size, morphological characterization reveals the homogeneous diameter of the fibers within the drug release platform. In Figure 1a–d, typical SEM images show a fibrillar structure with diameters up to 0.6 μm. The obtained average fiber diameters and the standard deviation are 0.42 ± 0.05 μm for EMFs-CUR, 0.52 ± 0.06 μm for EMFs-MSNs@CUR, 0.60 ± 0.08 μm for EMFs-KET, and 0.54 ± 0.08 μm for EMFs-MSNs@KET. Despite the standard deviation values representing less than 15% in all cases, it is evident that the values are higher in the presence of KET. This effect can be associated with modifications in the viscoelastic precursor solution properties induced by the insolubility of KET in organic solvents. The heating during the precursor solution elaboration improves the solubility of KET and the presence of the other components allows the integration of the drug in the system. The entanglement of polymeric chains during the solution's elaboration can host the KET crystals, making them available for the release. On the other hand, CUR is completely soluble in organic solvents, allowing the complete integration of the molecules in the polymeric matrix and thus resulting in a lower average diameter and standard deviation.

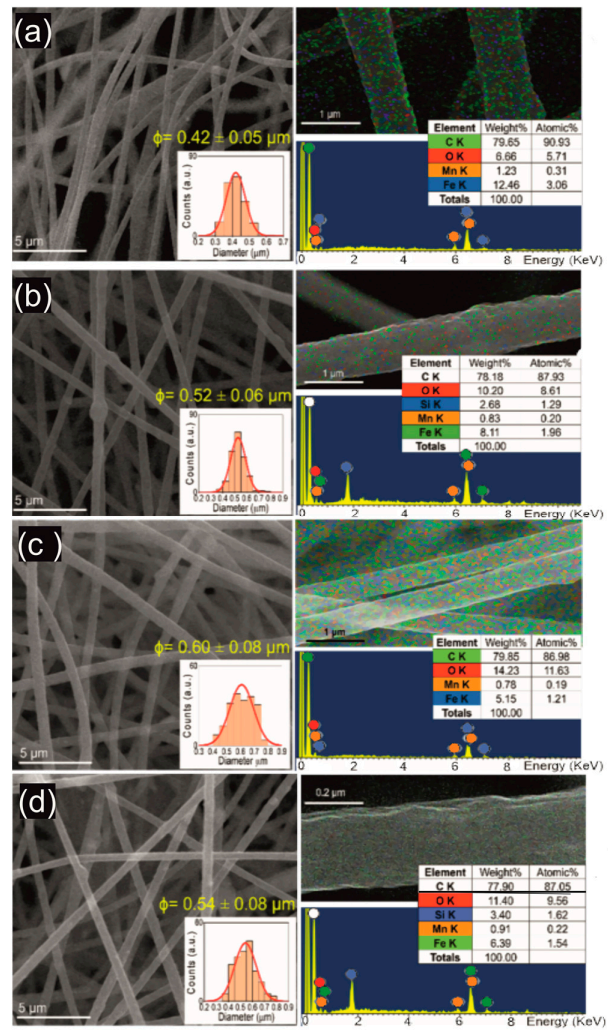

**Figure 1.** Morphology and composition of elaborated EMFs as drug release platform. Typical electron micrographs showing homogeneous fibrillar configuration in all samples (**a**–**d**), with diameters up to 0.6 µm. Furthermore, EDS analysis reveals adequate encapsulation of magnetic nanoparticles in EMFs-CUR (**a**) and EMFs-KET (**b**) samples, while it is also possible to appreciate the mesoporous silica nanostructures for EMFs-MSNs@CUR (**c**) and EMFs-MSNs@KET (**d**).

According to the EDS survey, the EMFs are mainly composed of C, Fe, Mn, and O elements, and in the cases of EMFs-MSNs@CUR and EMFs-MSNs@KET, Si interactions were detected. Taking into account in the estimation that the composition of MNPs is $Fe_{2.9}Mn_{0.1}O_4$ (see Section S1.2 in Supplementary Materials), EMFs-CUR presents the higher yield of encapsulation of MNPs with 14% in weight, while EMFs-KET only reaches 6%. However, regarding the Si contents, EMFs-MSNs@KET shows 3.40% MSNs encapsulation and EMFs-MSNs@CUR only 2.68%. In the presence of MSNs, the MNPs encapsulation capacity is also decreased, reaching 9% in EMFs-MSNs@CUR and 7.20% for EMFs-MSNs@KET. The homogeneous distribution of the elements can be observed in the mapping images (Figure 1), suggesting that the MNPs and MSNs are well-distributed along the fibers and that the developed elaboration methods are adequate for the drug release platform's elaboration with controlled morphology and composition.

To elucidate the possible modifications during the EMFs' processing, the FTIR analysis monitored the interactions between the IR radiation and the functional groups (normal modes of vibration). In Figure 2, the FTIR spectra from the used drugs were blank, and the corresponding results for the different EMFs samples are presented to observe the spectral modifications as a result of the materials processing. In case of EMFs-CUR and EMFs-KET samples, the absorbance of the representative vibrational modes of the CUR and KET molecules was (600–1600 cm$^{-1}$). The vibrations attributed predominantly to both molecules can be associated with alkenes (C=C) and carbonyl (C=O) (1628 cm$^{-1}$). Meanwhile, the stretching vibration at 3200–3500 cm$^{-1}$ was due to O-H groups; the C=C aromatic stretching vibration (1427 cm$^{-1}$) and 1512 cm$^{-1}$ were attributed to normal modes vibration, as were the stretching carbonyl ν(C=O), the in-plane bending around aliphatic δ CC-C, δ CC=O and the in-plane bending vibrations around the aromatic δ CC-H of keto and enol configurations from KET and CUR [35,36].

On the other hand, the normal modes from the PAN chains ((CH$_2$CHCN)$_n$) can be appreciated as masking to the drug molecule. The normal mode associated with $v_{as}$ −CH$_2$− was detected at 2930 cm$^{-1}$, and the characteristic C≡N was (2247 cm$^{-1}$) and for C=N it was (1233 cm$^{-1}$). At 1450 cm$^{-1}$, the mode is associated with CH$_2$ bending (scissoring) and at 1170 cm$^{-1}$ it is associated with the unsaturated esters and carboxylic acids [37]. In addition, for the case of EMFs-MSNs@CUR and EMFs-MSNs@KET, the silica interactions mask the polymeric and drug vibrational modes in the 600–1217 cm$^{-1}$ interval as reported previously [38]. This effect suggests that the interactions among the components hide the individual interactions and reduce the functional group's freedom degree to interact with radiation for normal vibrational modes detection. Based on these observations, it is probable that any chemical interactions among the components of EMFs occur during the materials processing. It suggests that the components do not suffer chemical modifications and are not chemically bonded. In a physical mixing, electrostatic, long range and Van der Waals interactions can be related to the system's configuration, making the drug available for its release in aqueous media.

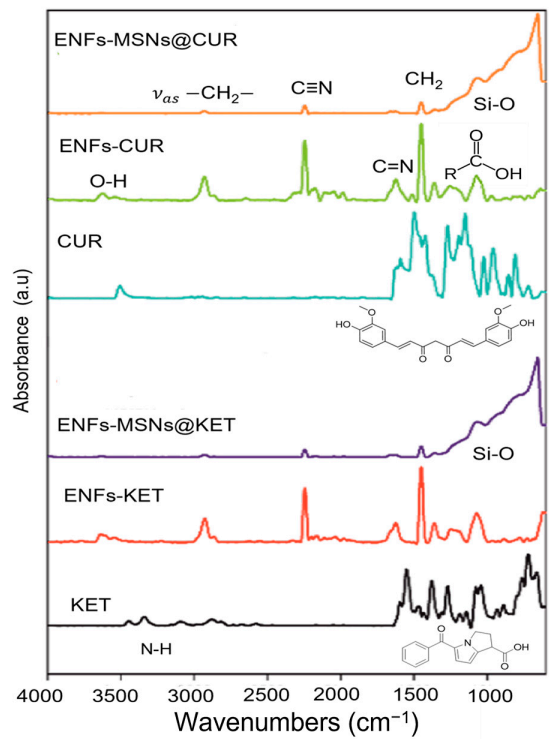

**Figure 2.** FTIR spectra from ketorolac and curcumin are compared with the different EMFs samples, showing that the interactions of the drugs are decreased by normal modes from polymer matrix and silica nanostructures.

PAN fibers have demonstrated thermal stability at least up to 300 °C and the observed weight loss below this temperature which can be associated with the thermal transitions of the drugs, at 180 °C for CUR and 150 °C for KET (Figure 3). Therefore, it is possible to estimate the amount of KET and CUR using TGA analysis, calculating the weight loss before the main transition at 322 °C. According to the observed results, the amount of CUR in EMFs- CUR is about 2.60 %wt., while for EMFs-KET, the amount of the drug is 2.40 %wt. In the case of EMFs-MSNs@CUR, only the 2 %wt. was encapsulated and only 1.30 %wt. in EMFs-MSNs@KET can be considered KET. For the transition of 322 and 329 °C, the nitrile functional groups from the PAN linear structure changed to a reticulated C=C structure. Usually, this transition occurs around 300 °C; however, because the mixing of the different components, this transition can experience a slight shift to higher temperatures.

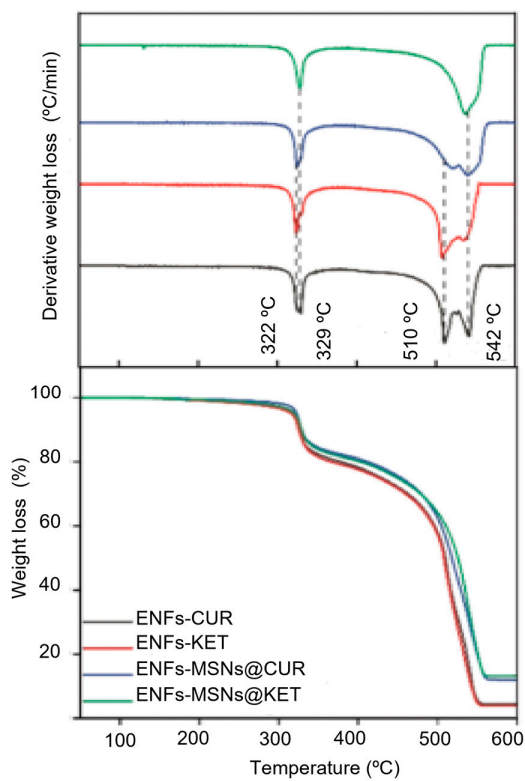

**Figure 3.** Thermogravimetric analysis results where the weight losses are associated mainly to the polymer interactions. However, it is important to notice shifts in the thermal transitions associated with the strong interactions among the components in the fibers. The residua can be considered inorganic nanoparticles contents and small contents of ashes; in the case of EMFs-CUR and EMFs-KET, up to 4 %wt. Meanwhile, for EMFs-MSNs@CUR and EMFs-MSNs@KET, it is up to 13 %wt.

The last transitions centered at 510 and 542 °C correspond to the pyrolysis of the polymer, losing almost all the weight in the samples. The remaining material after 550 °C can be considered as ashes and inorganic nanoparticles in all cases. The rest of the material associated with MNPs in the EMFs-CUR samples was 4.5 %wt, while for EMFs-KET, it was a lower amount, approximately 3.9 %wt by weight of the MNPs content. On the other hand, for the EMFs-MSNs@CUR samples, the remaining 12 %wt. and 13.2 %wt. for EMFs-MSNs@KET are associated with a combination of MNPs and MSNs embedded in the fibers.

These results agree with the EDS analysis, where it is possible to observe improved MNPs encapsulation using CUR, and also in the case of KET, where the MSNs contents is improved. The amount of available drug in EMFs-CUR and KET was determined through dissolving the fibers in DMF and comparing with a calibration curve in both the organic solvent and in the presence of a polymer. The drug-loading capacity of EMFs-MSNs@CUR and EMFs-MSNs@KET was estimated from the loading capacity of the MSNs and the contents of MSNs in the fibers. The maximum loading capacity in each case is 0.085 $\frac{\mu mol}{mg}$ for EMFs-CUR, 0.022 $\frac{\mu mol}{mg}$ for EMFs-KET, 0.035 $\frac{\mu mol}{mg}$ EMFs-MSNs@CUR, and 0.015 $\frac{\mu mol}{mg}$ EMFs-MSNs@KET.

The magnetic properties of the EMFs-MSNs@KET and EMFs-MSNs@CUR samples were evaluated at 300 K as a representative model to observe the effects of embedding the MNPs, together with therapeutic molecules and diamagnetic MSN. Magnetization curves, shown in Figure 4a,b, illustrate that both samples exhibit a magnetic saturation ($M_s$) of approximately 6 emu/g. The coercive field ($H_c$), which is the applied magnetic field required to reduce the magnetization to zero, is lower in the EMFs-MSNs@KET sample (53 Oe) compared to the EMFs-MSNs@CUR sample (75 Oe). The saturation magnetization ($M_s$) of the MNPs was found to be $M_s$ = 54 emu/g, with a coercive field of $H_c$ = 12 Oe. The corresponding $M_s$ value of the EMFs was approximately nine times lower than the MNPs, which was higher than expected since ≈80% of the EMFs' mass that corresponds to the nonmagnetic polymer matrix would give $M_s$ values of 9–11 emu/g.

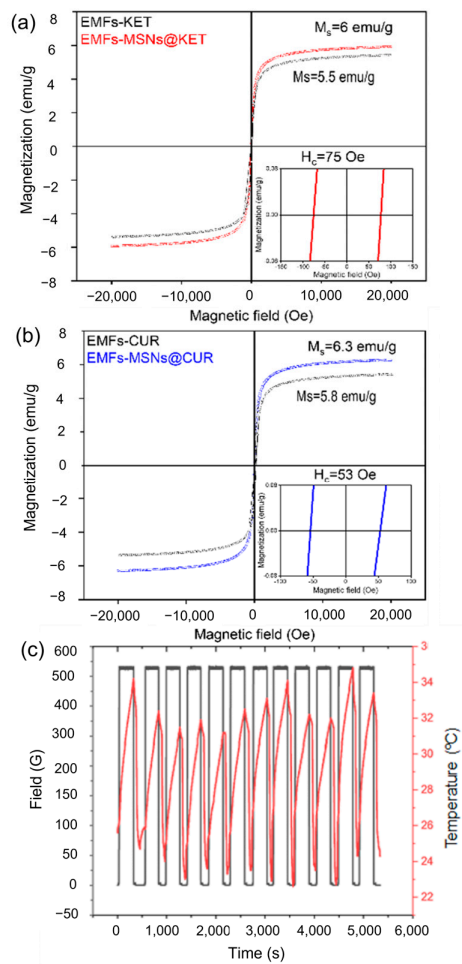

**Figure 4.** Magnetic responses from the elaborated EMFs. Magnetization curves of (**a**) EMFs-CUR and EMFs-MSNs@CUR and (**b**) EMFs-KET and EMFs-MSNs@KET. The fibers show a magnetic saturation of 6 and 6.3 emu/g, respectively, which is about 9 times lower than that of the MNPs used

(54 emu/g), suggesting 1/9 of the mass of the fibers is composed by magnetic materials. This value is reduced in the case of EMFs-MSNs samples. Heating dissipation response applying AMF on EMFs-MSNs@CUR samples (**c**). Cycles of 300 s AMF-on and 300 s AMF-off were evaluated 12 times to ensure repeatability and stability on the heating responses. Inductive heating dissipation is capable of increasing the liquid s temperature up to 12 °C.

$M_s$ is an intrinsic property determined by the type and arrangement of the atoms and their interactions among magnetic moments. It represents the maximum magnetization that the EMFs can achieve when all magnetic moments are aligned under the stimulation of an applied magnetic field. The $M_s$ values are primarily influenced by the number of unpaired electrons within the MNPs structure and the exchange interactions between them. In this case, when the MNPs are mixed with non-magnetic material (PAN) or diamagnetic (MSNs), a saturation magnetization decrease is expected. The effect of the non-magnetic material diluting the magnetic moments in the MNFs and the reduced values of $M_s$ were observed.

The overall magnetic properties of the MNFs depend on the concentration, distribution, and interaction of the MNPs within the fibrillar matrix. Together with the larger $H_C$ values observed in the EMFs, it is plausible that the partial agglomeration of the MNPs during the electrospinning fabrication results in a measurable magnetic interaction between the particles, a decrease in MS and the magnetic hardening of the composite material. The observed M vs. H hysteresis loops measured at room temperature showed saturation magnetization values between 5 and 7 emu/g of the EMFs and very small coercive fields, as expected from the constituting magnetic material.

Moreover, the embedded MNPs showed SLP values up to 85 W/g at 765 kHz and 364 G (Section S1.2 in Supporting Materials), ensuring magnetic heating response. The inductive heating of the EMFs in a liquid carrier (Figure 4c) showed heating rates up to 0.05 K/s. Samples from EMFs-MSNs@CUR were selected for the evaluation of heating dissipation under the stimulation of 30 s pulses-on and 30 s pulses-off stimulation.

The application of heat to the skin can enhance drug permeation through the skin's barrier and increase the diffusion rate of drugs into the underlying tissues. Heat can disrupt the structure of the stratum corneum, the outermost layer of the skin, by increasing its permeability. This allows for better penetration of drugs through the skin barrier. Heat-induced vasodilation also increases blood flow to the skin, facilitating drug absorption. Moreover, heating can increase the solubility and mobility of drugs in the skin and the underlying tissues, promoting their dissolution and facilitating their diffusion. This is particularly beneficial for drugs with limited solubility or those that require higher temperatures to dissolve effectively. It is important to highlight the fact that the application of heat should be carefully controlled to ensure safety and avoid skin damage. The temperature and duration of heating should be optimized to achieve the desired enhancement in drug release without causing harm to the skin or compromising the stability of the drug.

The heat generated by the EMFs within 300 s is sufficient to raise the temperature of the circulating liquid medium by up to 12 °C, starting from temperatures under 25 °C, proving possible the EMFs ability to reach the target temperatures 37–45 °C that enhance drug permeation through the skin via localized heating.

*3.2. Drug Release Platform Responses*

The spontaneous drug release from EMFs was evaluated (field OFF) and after in ON conditions, where the magnetic stimulation was applied (765 kHz, 364 G).

3.2.1. Alternating Magnetic Field OFF

The strategy to evaluate the EMFs as a drug release platform was started without any applied AMF, establishing a baseline for the comparison of spontaneous release against the triggered response. Samples of each different type of EMFs were incubated at 37 °C under agitation for a total period of ten days. The spontaneous release profiles from EMFs-CUR, EMFs-KET, EMFs-MSNs@CUR, and EMFs-MSNs@KET at different pH (7.4 and 5.5) simulating neutral biologic media (phosphate buffer) and acidic conditions (acetate buffer) mimicking a healthy skin environment, are shown in Figure 3. EMFs-KET released almost all the loaded drugs within 10 days, exhibiting a noticeable difference depending on the pH value, with a maximum release of 94.4% at pH 7.4 and 85.7% at pH 5.5, respectively, of the available drug (Figure 5a). This trend is similar to the release from EMFs-MSNs@ KET and is consistent with the lower solubility of KET at a lower pH. On the other hand, EMFs-CUR showed a much lower release with a maximum of 3% after 10 days, consistent with the low solubility of CUR in aqueous media. Noticeably, the pH value has almost no effect on the release of CUR from the platform.

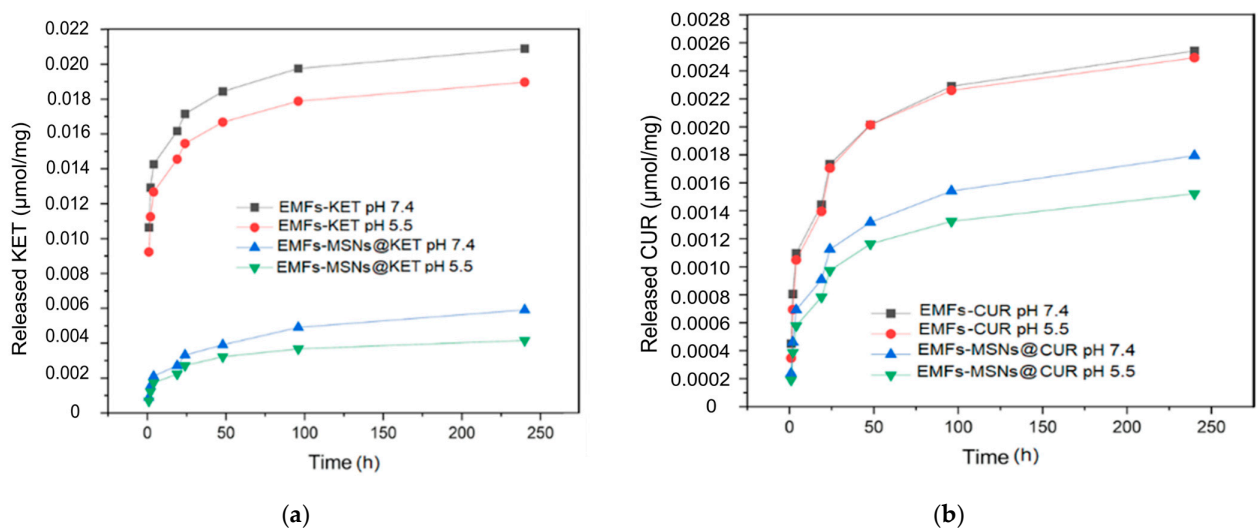

(**a**)　　　　　　　　　　　　　(**b**)

**Figure 5.** Spontaneous drug-release from EMFs platform. (**a**) EMF-KET and EMF-MSNs@KET and (**b**) EMF-CUR and EMF-MSNs@CUR evaluated in pH 7.4 (phosphate buffer) and 5.5 (acetate buffer) mimicking neutral pH and healthy skin conditions, respectively.

A direct comparison between the release rates from the platform with the free drug and the one with the drug encapsulated in mesoporous silica nanoparticles (MSNs) shows release patterns, as shown in Figure 3. The release rate from EMFs without MSNs is notably higher, indicating that the extra layer of encapsulation provided by the MSNs reduces the system's drug release rate. In all instances, except for the EMFs containing ketorolac (EMFs-KET), less than 90% of the drug loaded into the EMFs was released. It is likely that in the case of EMFs-KET, the drug is securely encapsulated by both the EMFs and MSNs, creating dual barriers to cross before release into the aqueous medium. Specifically, for curcumin (CUR), its affinity for organic solvents could lead to strong interactions with the polymer chains, hindering the drug molecules from being released.

3.2.2. Alternating Magnetic Field ON

Under the applied field ($t_{ON}$ = 5 min; f = 765 kHz; $H_0$ = 364 G), the release of ketorolac tromethamine increased significantly (Figure 6a,b). For EMFs-KET, the amount released within 1 h increased to 85%. While this is already a very significant increase, the effect of the magnetic release stimulus is much higher in EMFs-MSNs@KET. In this sample, the release after 1 h increased to 0.021 μmol/mg, showing an enhanced release from the platform controlled by the inductive heating. As discussed before, the encapsulation of the drugs inside MSNs offers slow-release rates. This property of the platform opens the gates for triggered drug release applications, with a minimum release before the stimulus is applied, when the maximum release is reached.

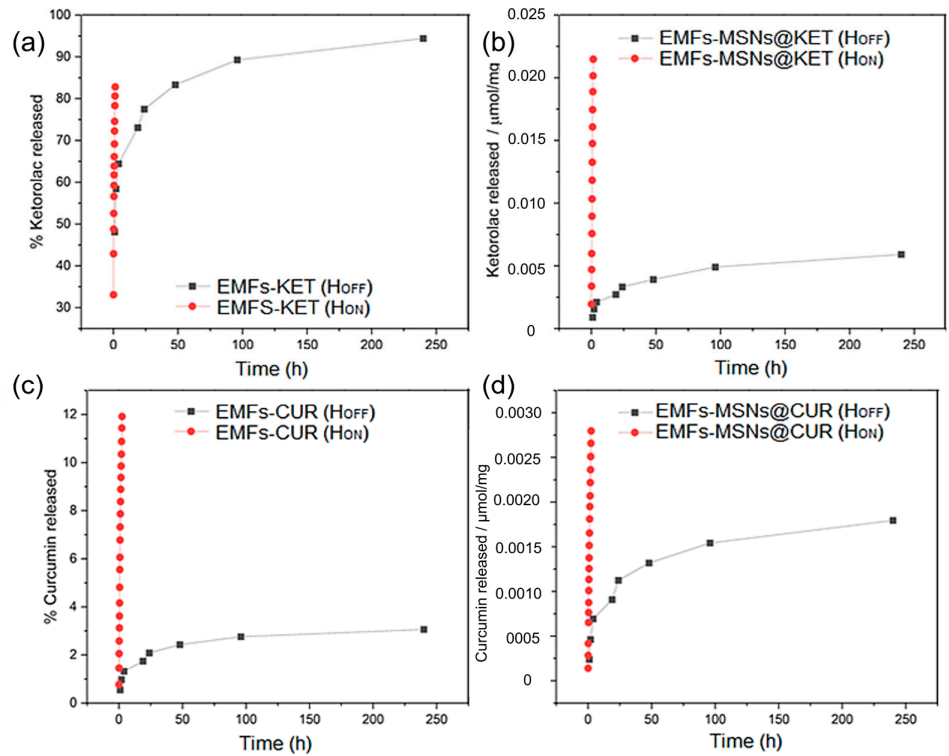

**Figure 6.** Comparison of the spontaneous release response from the EMFs platform (H$_{OFF}$) with the magnetically triggered release response (H$_{ON}$) of EMFs-KET (**a**), EMFs-MSNs@KET (**b**), EMFs-CUR (**c**) and EMFs-MSNs@CUR (**d**). The effect of the alternating magnetic field application (765 kHz, 364 G) in pulses of 5 min during 120 min improves the amount of released drug compared with the spontaneous release at pH 7.4.

In the case of the low-water soluble CUR, the increase in the amount of released drug is more evident through the magnetic stimulation in Figure 4c,d. The spontaneous release only reached 3%; meanwhile, using magnetic stimulation, up to 12% of the available CUR can be released. This improvement has not only has an impact in the amount of CUR delivered, it also represents the possibility to remotely control the time schedules for drug administration using AMF that are non-harmful.

The application of AMF allows an increase in the temperature, and therefore the solubility of both molecules is improved. CUR s solubility in water is temperature dependent; is it possible to achieve a 12-fold increase at 90 °C with respect to room temperature [39–41]. Meanwhile for KET, despite the drug being soluble in water, the bioavailability and availability can be improved using heat, pH and other components in the dosage form [42–44].

For the case of EMFs-CUR and EMFs-CUR, both the drugs are combined with the MNFs and embedded within EMFs; thus, the availability is increased with respect to EMFs-MSNs@CUR and EMFs-MSNs@KET. After the AFM stimuli, the release from the EMFs-DRUG is higher than the EMFs-MSNs@DRUG, because of the diffusion barrier that the porous structure of the MSNs creates. This complex pathway slows down the release rate compared to a straightforward diffusion in a homogeneous medium.

The reduced release of CUR and KET loaded into MSNs is primarily reduced by the controlled pore size, surface chemistry, and the physical structure of the MSNs. The CUR and KET molecules must be released from an interconnected network of pores, which acts as a physical barrier to rapid release. In this case, the MSNs collectively create a barrier against spontaneous release and rapid diffusion. The combined mechanisms (double layer release) from MSNs and the fibers allow for a sustained and targeted release of the encapsulated molecules triggered by heating, making MSNs a versatile platform for drug delivery and other applications requiring a controlled release of soluble and non-soluble therapeutic molecules.

## 4. Conclusions

The drug-loaded magnetically responsive material studied here, based on electrospun nanofibers with MNPs, represents a functional system designed to overcome the hurdles of administering poorly soluble or insoluble drugs in water-based media, for instance biological environments. The elaborated platform can be employed for hydrophilic and hydrophobic drug, using KET as a representative and CUR as a model drug in the evaluation. For both molecules, the spontaneous drug release profiles were determined and the observations showed that when the molecules are integrated in the MNFs, a reduced amount of CUR is released compared with the amount of KET released. When both molecules are loaded in MSNs, a lower amount is released.

The produced EMFs possess the capacity for inductive heating and the ability to heat up the surrounding medium, with 1.3 mg of MNFs being possible to produce heating in 1 mL of PBS up to 8 °C within a 5 min AMF-on pulse. The MNFs showed an overall specific loss power of 55 W/g. While these measurements were performed in the release medium (PBS or acetate buffer), the fixation of the magnetic particles in the polymer fibers should also prevent any contributions from Brownian relaxation. The heating values confirm that the EMFs can still provide hyperthermia in addition to stimulated drug release, therefore constructing new platforms for the delivery of soluble and non-soluble drugs remotely controlled by harmless magnetic fields. These results constitute a clear way to control the drug release profile and kinetics, offering potential advantages in applications such tissue engineering and wound healing formulations where sustained release is required.

**Acknowledgments:** Access to the facilities of the Servicio General de Apoyo a la Investigación-SAI, Universidad de Zaragoza is acknowledged. The authors thank the University of Cologne for infrastructural and financial support in the framework of the UoC Forum, iRNA-Carriers "Transformative Nanocarriers for RNA Transport and Tracking".

**Conflicts of Interest:** The authors declare no conflicts of interest.